\documentclass[twocolumn,prl,superscriptaddress,showpacs,amsmath,amssymb,10pt]{revtex4}

\usepackage{graphicx}
\usepackage{dcolumn}
\usepackage{stmaryrd}

\begin{document}

\title{{\normalsize Macroscopic Kinetic Effect of Cell-to-Cell Variation in Biochemical Reactions}}

\author{Pan-Jun Kim}
\affiliation{Institute for Genomic Biology, University of Illinois, Urbana, Illinois 61801, USA}
\author{Nathan D. Price} \email{ndprice@illinois.edu}
\affiliation{Institute for Genomic Biology, University of Illinois, Urbana, Illinois 61801, USA}
\affiliation{Department of Chemical and Biomolecular Engineering and Center for Biophysics and Computational Biology, University of Illinois, Urbana, Illinois 61801, USA}


\begin{abstract}

Genetically identical cells under the same environmental conditions can show strong variations in protein copy numbers due to inherently stochastic events in individual cells. We here develop a theoretical framework to address how variations in enzyme abundance affect the collective kinetics of metabolic reactions observed within a population of cells. Kinetic parameters measured at the cell population level are shown to be systematically deviated from those of single cells, even within populations of homogeneous parameters. Because of these considerations, Michaelis-Menten kinetics can even be inappropriate to apply at the population level. Our findings elucidate a novel origin of discrepancy between {\it in vivo} and {\it in vitro} kinetics, and offer potential utility for analysis of single-cell metabolomic data.

\end{abstract}

 \maketitle
Noisy or stochastic molecular events are plentiful in the life of a cell. In the past several years, extensive experimental and theoretical efforts have been devoted to analyzing stochastic processes in gene and protein expression, at both the transcriptional and translational levels \cite{rsto}. One consequence of stochastic gene expression is that the number of molecules of a given protein can vary substantially from cell to cell, even within genetically identical populations \cite{rlog1,rlog2}. Such stochastic gene expression has received considerable attention in relation to cellular regulation, phenotypic diversity, and disease \cite{rsto,rlog1,rlog2,rpheno}.

Kinetic modeling of metabolism aims at achieving a quantitative description of biochemical reactions to generate mass and energy required for cell survival. Since there are rarely available techniques to detect metabolites inside a single cell \cite{rsglmet}, most kinetic models for intracellular metabolism have been built on experimental data obtained from cell populations. These models, however, would be valid when behavior of individual cells is very similar to the average behavior of the population. Significant variation of enzyme abundance between cells might challenge this traditional approach, and here we investigate how single-cell variation affects the kinetics of metabolic reactions appearing at the population level. 

The flux of any given reaction ($\nu$) can be expressed as a function of the molecular concentrations and the kinetic constants associated with the reaction: $\nu = f(E,\{S_{\gamma}\},\{K_{i}\})$, where $E$ stands for the concentration of enzyme, $S_{\gamma}$ for the concentration of the $\gamma$th metabolite participating in or allosterically regulating the reaction, and $K_{i}$ for the $i$th kinetic constant in the reaction. For example, if a single-substrate reaction follows the Michaelis-Menten (MM) kinetics, then $\nu = f(E,S,\{K_{o},K_{M}\}) = K_{o}ES/(K_{M}+S)$. Usually, experiments measure concentrations or fluxes {\it averaged} over a cell population, as appears below.
\begin{align}\label{grd1}
\langle\nu_{z}\rangle_{z}&=\langle f(E_{z},\{S_{\gamma z}\},\{K_{i}\})\rangle_{z}\nonumber\\
&= f(\langle E_{z}\rangle_{z},\{\langle S_{\gamma z}\rangle_{z}\},\{K_{i}\})\times (1+\xi)\,,
\end{align}
where we index each cell with $z$ such that $\nu_{z}$ is the reaction flux inside cell $z$, and $\langle\cdots\rangle_{z}$ averages a value over a cell population. The last equality gives the definition of $\xi$ to be equal to zero if molecules are uniformly distributed over the population. For clarity of analysis, here we only consider cellular variability of molecular concentrations but not of kinetic parameters \cite{SXie} and cell volumes; in Eq. (\ref{grd1}), there is no $z$-dependency of $K_{i}$ and no weighting of cell volumes for the averages. Although this analysis can be extended straightforwardly for the factors excluded here, our work demonstrates that experimentally well-established heterogeneity in single-cell enzyme abundance itself \cite{rlog2,rdata} gives rise to inevitable and definite effects on the kinetic properties of metabolic reactions. In the following, we consider the steady states of reactions under a given enzyme distribution, as the typical time scale of reaction rates is much shorter ($\sim 10^{-3}\rm s$) than that of enzyme concentration changes ($>10^{2}\rm s$) \cite{Stryer}.

Applying the series expansion, we can write $\xi$ as
\begin{align}\label{grd2}
\xi =&\frac{1}{f_{0}}\Bigg\{\frac{1}{2}\cdot\frac{\partial^{2}f}{\partial E^{2}}\Big|_{0}\langle(\Delta E_{z})^{2}\rangle_{z}+\sum_{\gamma}\frac{\partial^{2}f}{\partial E\partial S_{\gamma}}\Big|_{0}\langle\Delta E_{z}\Delta S_{\gamma z}\rangle_{z}\nonumber\\
&+\frac{1}{2}\sum_{\gamma}\sum_{\gamma'}\frac{\partial^{2}f}{\partial S_{\gamma}\partial S_{\gamma'}}\Big|_{0}\langle\Delta S_{\gamma z}\Delta S_{\gamma' z}\rangle_{z}+\cdots\Bigg\}\,,
\end{align}
where the higher-order terms have been omitted. $f_{0}\equiv f(\langle E_{z}\rangle_{z},\{\langle S_{\gamma z}\rangle_{z}\},\{K_{i}\})$, $f\equiv f(E,\{S_{\gamma}\},\{K_{i}\})$, $\Delta E_{z}\equiv E_{z}-\langle E_{z}\rangle_{z}$, $\Delta S_{\gamma z}\equiv S_{\gamma z}-\langle S_{\gamma z}\rangle_{z}$, and $|_{0}$ is for derivative around $E=\langle E_{z}\rangle_{z}$, $S_{\gamma}=\langle S_{\gamma z}\rangle_{z}$. Note that (i) if $f$ is a simple linear function of molecular concentrations, $\xi$ vanishes, and (ii) variability of the concentrations across cells explicitly contributes to $\xi$. In other words, inherent nonlinearity in reaction kinetics and significant cellular variation in molecular levels manifest the difference between $\langle\nu_{z}\rangle_{z}$ and $f(\langle E_{z}\rangle_{z},\{\langle S_{\gamma z}\rangle_{z}\},\{K_{i}\})$. Hence, kinetic constants calculated from a population do not necessarily coincide with $\{K_{i}\}$ at a single-cell level, as will be addressed in detail.

Suppose that $\langle\nu_{z}\rangle_{z}$, $\langle E_{z}\rangle_{z}$, $\{\langle S_{\gamma z}\rangle_{z}\}$ are experimentally measured to achieve kinetic constants $\{\widetilde{K}_{i}\}$ fitting $f(\langle E_{z}\rangle_{z},\{\langle S_{\gamma z}\rangle_{z}\},\{\widetilde{K}_{i}\})$ to $\langle\nu_{z}\rangle_{z}$, for example, by minimizing the error $\epsilon = \left|\left(f(\langle E_{z}\rangle_{z},\{\langle S_{\gamma z}\rangle_{z}\},\{\widetilde{K}_{i}\})-\langle\nu_{z}\rangle_{z}\right)/\langle\nu_{z}\rangle_{z}\right|$. This is the typical approach. If more than one experimental dataset are available, $\{\widetilde{K}_{i}\}$ may instead be obtained by minimizing $\sum_{r} \epsilon^{2}_{r}$, where subscript $r$ represents the $r$th experiment. Combined with Eq. (\ref{grd1}), this procedure results in the following formula for $\{\widetilde{K}_{i}\}$ at the lowest-order approximation:
\begin{equation}\label{gen1}
\rm{\Delta\bold{K}\approx\bold{A^{-1}u}}\,,
\end{equation}
where $\rm{\Delta\bold K}$ and $\rm\bold u$ are the vectors whose elements are
\begin{displaymath}
\Delta K_{i}=\widetilde{K}_{i}-K_{i}\,,\,\,\,u_{i}=\sum_{r}\frac{\xi_{r}}{f_{0r}}\frac{\partial f}{\partial K_{i}}\Big|_{0r}\,,
\end{displaymath}
respectively, and $\rm\bold A^{-1}$ is the inverse matrix of $\rm\bold A$ with elements
\begin{displaymath}
A_{ij}=\sum_{r}\frac{1}{f^{2}_{0r}}\frac{\partial f}{\partial K_{i}}\Big|_{0r}\frac{\partial f}{\partial K_{j}}\Big|_{0r}\,.
\end{displaymath}
It should be noticed that Eq. (\ref{gen1}) gives a degree of discrepancy between kinetic constants at a population level $\{\widetilde{K}_{i}\}$ and those at a single-cell level $\{K_{i}\}$, as a function of a degree of cellular heterogeneity in Eq. (\ref{grd2}).

Taking into account a single-substrate reaction governed by the MM kinetics [$f=K_{o}ES/(K_{M}+S)$], it is straightforward to get $\widetilde{K}_{o}$ and $\widetilde{K}_{M}$ from Eq. (\ref{gen1}) if both come out of the same population data:
\begin{align}\label{MM1}
&\frac{\Delta K_{o}}{K_{o}}\approx\langle\xi_{r}\rangle_{r}-\frac{{\rm cov}_{r}\left(\xi_{r},\frac{1}{1+\langle S_{zr}\rangle_{z}/K_{M}}\right)}{{\rm var}_{r}\left(\frac{1}{1+\langle S_{zr}\rangle_{z}/K_{M}}\right)}\left\langle\frac{1}{1+\frac{\langle S_{zr}\rangle_{z}}{K_{M}}}\right\rangle_{r}\,,\nonumber\\
&\frac{\Delta K_{M}}{K_{M}}\approx-\frac{{\rm cov}_{r}\left(\xi_{r},\frac{1}{1+\langle S_{zr}\rangle_{z}/K_{M}}\right)}{{\rm var}_{r}\left(\frac{1}{1+\langle S_{zr}\rangle_{z}/K_{M}}\right)}\,,
\end{align}
where ${\rm cov}_{r}(x_{r},y_{r})\equiv \langle x_{r}y_{r}\rangle_{r}-\langle x_{r}\rangle_{r}\langle y_{r}\rangle_{r}$, ${\rm var}_{r}(x_{r})\equiv \langle x^{2}_{r}\rangle_{r}-\langle x_{r}\rangle^{2}_{r}$. From Eq. (\ref{MM1}), one can further prove the following relation:
\begin{equation}\label{MM2}
\sqrt{\left(\frac{\Delta K_{o}}{K_{o}}\right)^{2}+\left(\frac{\Delta K_{M}}{K_{M}}\right)^{2}}\gtrsim\frac{|\langle\xi_{r}\rangle_{r}|}{\sqrt{1+\left\langle\frac{1}{1+\langle S_{zr}\rangle_{z}/K_{M}}\right\rangle^{2}_{r}}}\,,
\end{equation}
which shows that the degree of deviations in kinetic parameters is essentially determined by $\langle\xi_{r}\rangle_{r}$, as the right side ranges from $|\langle\xi_{r}\rangle_{r}|/\sqrt{2}$ to $|\langle\xi_{r}\rangle_{r}|$. On the other hand, if $K_{M}$ is known at the single-cell level, one might calculate only $\widetilde{K}_{o}$ from experimental data, but not $\widetilde{K}_{M}$. In this case, the following simplified relation from Eq. (\ref{gen1}) holds:
\begin{equation}\label{MM3}
\frac{\Delta K_{o}}{K_{o}}\approx\langle\xi_{r}\rangle_{r}\,.
\end{equation}
Since $\xi_{r}$s consistently include the negative-sign terms (as shown below), they would not be simply canceled out by each other under averaging, thereby allowing for significant nonzero $\langle\xi_{r}\rangle_{r}$ in Eqs. (\ref{MM2}) and (\ref{MM3}). For simplicity of analysis, we will concentrate upon cases of Eq. (\ref{MM3}) out of single experiments. Note that $S_{z}=K_{M}\nu_{z}/(K_{o}E_{z}-\nu_{z})$; thus, $\Delta S_{z}$ in Eq. (\ref{grd2}) can be substituted for by $\Delta E_{z}$ and $\Delta\nu_{z}\equiv\nu_{z}-\langle\nu_{z}\rangle_{z}$ to give,
\begin{align}\label{MMx}
\xi\approx&-\left(1+\frac{\langle S_{z}\rangle_{z}}{K_{M}}\right)\frac{\langle(\Delta E_{z})^{2}\rangle_{z}}{\langle E_{z}\rangle^{2}_{z}}-\frac{\langle S_{z}\rangle_{z}}{K_{M}}\cdot\frac{\langle(\Delta\nu_{z})^{2}\rangle_{z}}{\langle\nu_{z}\rangle^{2}_{z}}\nonumber\\
&+\left(1+2\frac{\langle S_{z}\rangle_{z}}{K_{M}}\right)\frac{\langle\Delta E_{z}\Delta\nu_{z}\rangle_{z}}{\langle E_{z}\rangle_{z}\langle\nu_{z}\rangle_{z}}\,.
\end{align}
As long as there exists variation in enzyme concentration, the first term on the right side of Eq. (\ref{MMx}) has a nonzero magnitude, always greater than $\langle(\Delta E_{z})^{2}\rangle_{z}/\langle E_{z}\rangle^{2}_{z}$ of which experimental values are recently available for the yeast {\it Saccharomyces cerevisiae} \cite{rdata}. Let $\xi_{A}$ be this first term, and we can estimate $\xi$ to be $\xi_{A}$ if $\langle(\Delta\nu_{z})^{2}\rangle_{z}$ is sufficiently small. Without the lowest-order approximation from Eq. (\ref{grd2}), we can also get $\xi$ from the exact formula by setting $\langle(\Delta\nu_{z})^{2}\rangle_{z}=0$, provided that the enzyme concentration follows the log-normal ($\equiv\xi_{L}$) or normal ($\equiv\xi_{N}$) distribution \cite{rlog1,rlog2}. Such $\xi_{L(N)}$ satisfies the following equality:
\begin{equation}\label{exact}
\frac{\langle S_{z}\rangle_{z}}{K_{M}}=\int_{1+\delta}^{\infty}\frac{\hat{P}_{L(N)}(x)}{x-1}dx\,,
\end{equation}
where $\hat{P}_{L(N)}(x)=(\langle\nu_{z}\rangle_{z}/K_{o})\times P_{L(N)}(E)$, $x=(K_{o}/\langle\nu_{z}\rangle_{z})\times E$, $\langle\nu_{z}\rangle_{z}=K_{o}\langle E_{z} \rangle_{z}\langle S_{z} \rangle_{z}/(K_{M}+\langle S_{z} \rangle_{z})\times (1+\xi_{L(N)})$, and $P_{L(N)}(E)$ is the probability distribution of enzyme concentration $E$ almost following the log-normal (normal) distribution. Because $P_{L(N)}(E)$ is approximated as the log-normal (normal) distribution, $\hat{P}_{L(N)}(x)$ can be approximated as the same, with $\langle x\rangle =(1+K_{M}/\langle S_{z}\rangle_{z})/(1+\xi_{L(N)})$ and $\sqrt{\langle(\Delta x)^{2}\rangle}/\langle x\rangle =\sqrt{\langle(\Delta E_{z})^{2}\rangle_{z}}/\langle E_{z}\rangle_{z}$ where $\Delta x\equiv x-\langle x\rangle$. $\delta >0$ is chosen small enough to satisfy $\int_{0(-\infty)}^{1+\delta}\hat{P}_{L(N)}(x)dx\ll 1$. Although $\xi_{L(N)}$ can be accurate for the particular forms of enzyme distribution following the log-normal (normal) distribution and $\xi_{A}$ is just an estimation, $\xi_{A}$ is potentially useful as can be applied without knowing a specific functional form of enzyme distribution.

We now turn our attention to the empirical values of $\xi_{A}$, $\xi_{L}$, and $\xi_{N}$. Table \ref{empirical} shows results for several irreversible reactions, which follow MM kinetics and have the necessary experimental data for calculating $\xi_{A}$, $\xi_{L}$, and $\xi_{N}$ for {\it S. cerevisiae}. It is then observed that $\xi_{A}$, $\xi_{L}$, and $\xi_{N}$ easily reach $\sim O(-10^{-1})$, thereby lowering $\widetilde{K}_{o}$ by several ten percent of $K_{o}$ according to Eq. (\ref{MM3}). Furthermore, a striking range of the metabolite concentration changes ($\sim 100$-fold increased or decreased) against different environmental conditions \cite{rmet} can overweight $\langle S_{z}\rangle_{z}/K_{M}$ in Eq. (\ref{MMx}) for certain conditions, which could drastically increase the magnitude of $\xi_{A}$ as well as of $\xi$.
\begin{table*}\renewcommand{\arraystretch}{1.4}
\caption{Characteristics of metabolic reactions and their $\xi_{A}$, $\xi_{L}$, and $\xi_{N}$ when $\delta =10^{-6}$. Marked $^{*}$ if a water molecule acts as a cosubstrate. Data for $\sqrt{\langle(\Delta E_{z})^{2}\rangle_{z}}/\langle E_{z}\rangle_{z}$ from Ref. \cite{rdata}, for $K_{M}$ from Ref. \cite{rdt1}, and for $\langle S_{z}\rangle_{z}$ from Ref. \cite{rdt2}.}
\begin{tabular}{ccccccccc}
\hline
\hline
Enzyme & Substrate & Product & $\sqrt{\langle(\Delta E_{z})^{2}\rangle_{z}}/\langle E_{z}\rangle_{z}$ & $K_{M}$ (mM) & $\langle S_{z}\rangle_{z}$ (mM) & $\xi_{A}$ & $\xi_{L}$ & $\xi_{N}$ \\
\hline
YEL042W$^{*}$ & GDP & GMP, phosphate & 0.235 & 0.1 & 0.39 & -0.271 & -0.260 & -0.278 \\
YJL005W & ATP & cAMP, diphosphate & 0.267 & 1.6 & 2.52 & -0.183 & -0.237 & -0.322 \\
YPL111W$^{*}$ & arginine & ornithine, urea & 0.241 & 15.7 & 50 & -0.244 & -0.261 & -0.291 \\
\hline
\hline
\end{tabular}
\label{empirical}
\end{table*}

\begin{figure}[h]
\begin{center}
\includegraphics[width=0.4\textwidth]{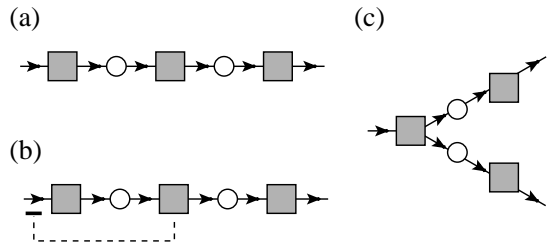}
\caption{Various types of metabolic pathways. Squares for metabolites and circles for reactions. Fluxes run into arrowed directions. (a) Linear, (b) linear with feedback inhibition depicted by a dashed line, (c) branching pathways.
}
\label{motif}
\end{center}
\end{figure}

When can $\xi_{A}$ be used to safely infer $\xi$? In a directed linear pathway as shown in Fig. \ref{motif}(a), each reaction takes as a substrate the product of the preceding reaction, and frequently involves an additional cosubstrate (such as a water molecule) that is abundant in the cell and whose variations can be neglected. For any given reaction in such a pathway, $\nu_{z}$ at a steady state is entirely determined by the influx from the upstream region independently of the given reaction itself ($\delta \nu_{z}/\delta E_{z}=0$), if there is absent any regulatory connection between the reaction and the upstream region. Thus the last term on the right side of Eq. (\ref{MMx}) vanishes because of decoupled $\nu_{z}$ and $E_{z}$, and a negative sign of the remaining second term even weights the effect of $\xi_{A}$ on $\xi$ because $\xi_{A}$ also has a negative sign. Therefore, the lower bound of the magnitude of $\xi$ in Eq. (\ref{MMx}) can be predicted by $\xi_{A}$.

If we consider more elaborate pathways than simple linear pathways, the last term on the right side of Eq. (\ref{MMx}) may not simply vanish, and could play a role in determining $\xi$. Specifically, if $E_{z}$ and $\nu_{z}$ are positively correlated, the last term will weaken the effect of $\xi_{A}$ on $\xi$, and if they are negatively correlated, will weight the effect in an opposite way. Here we focus on the former case, as examplified in Figs. \ref{motif}(b) and \ref{motif}(c). Figure \ref{motif}(b) depicts a negative-feedback case where the substrate of a given reaction inhibits the first reaction in the pathway or the transport of its precursor. In such a way, flux is reduced when the substrate is accumulated, as characterized by $\nu_z = c_{z}/\{1+(S_{z}/K_{I})^{h}\}$ where $c_{z}$ is the maximal influx, $K_{I}$ is the dissociation constant of the inhibiting interaction, and $h$ is a Hill coefficient. Substituting this into Eq. (\ref{MMx}) gives rise to,
\begin{align}\label{ih}
\xi\approx-(1-\alpha)\left\{1+\frac{\langle S_{z}\rangle_{z}}{K_{M}}(1-\alpha)\right\}\frac{\langle(\Delta E_{z})^{2}\rangle_{z}}{\langle E_{z}\rangle^{2}_{z}}+\cdots\,,
\end{align}
where $\alpha=[1+\{1+(K_{I}/\langle S_{z}\rangle_{z})^{h}\}h^{-1}(1+\langle S_{z}\rangle_{z}/K_{M})^{-1}]^{-1}$ \cite{ihco}. Compared with Eq. (\ref{MMx}), the contribution of the term with $\langle(\Delta E_{z})^{2}\rangle_{z}/\langle E_{z}\rangle^{2}_{z}$ is relatively small, and even approaches zero at a strong inhibition limit ($\langle S_{z}\rangle_{z}\gg K_{I}$, $h\gg 1$). This effect originates from the presence of a positive correlation between $E_{z}$ and $\nu_{z}$, as anticipated above. A similar effect can also be found from the case of Fig. \ref{motif}(c). In branching pathways, such as depicted in Fig. \ref{motif}(c), different enzymes can bind to the common substrate, and each catalyzes a first reaction in a different pathway. More specifically, a metabolite $S_{z}$ with influx $c_{z}$ is converted to a product with rate $\nu^{I}_{z}=K^{I}_{o}E^{I}_{z}S_{z}/(K^{I}_{M}+S_{z})$ or to the other product with rate $\nu^{II}_{z}=K^{II}_{o}E^{II}_{z}S_{z}/(K^{II}_{M}+S_{z})$. Equation (\ref{MMx}) then leads to,
\begin{align}\label{branch}
\xi^{I}\approx-\frac{\left(1+\frac{\langle S_{z}\rangle_{z}}{K^{I}_{M}}\right)\left(1+\frac{\langle\nu^{II}_{z}\rangle_{z}/\langle\nu^{I}_{z}\rangle_{z}}{1+\langle S_{z}\rangle_{z}/K^{II}_{M}}\right)}{\left\{1+\left(1+\frac{\langle S_{z}\rangle_{z}}{K^{I}_{M}}\right)\left(\frac{\langle\nu^{II}_{z}\rangle_{z}/\langle\nu^{I}_{z}\rangle_{z}}{1+\langle S_{z}\rangle_{z}/K^{II}_{M}}\right)\right\}^{2}}\cdot\frac{\langle(\Delta E^{I}_{z})^{2}\rangle_{z}}{\langle E^{I}_{z}\rangle^{2}_{z}}-\nonumber\\
\frac{\frac{1+\langle S_{z}\rangle_{z}/K^{I}_{M}}{1+\langle S_{z}\rangle_{z}/K^{II}_{M}}+\frac{\langle\nu^{I}_{z}\rangle_{z}}{\langle\nu^{II}_{z}\rangle_{z}}\cdot\left(1+2\frac{\langle S_{z}\rangle_{z}}{K^{I}_{M}}\right)}{\left(\frac{1+\langle S_{z}\rangle_{z}/K^{I}_{M}}{1+\langle S_{z}\rangle_{z}/K^{II}_{M}}+\frac{\langle\nu^{I}_{z}\rangle_{z}}{\langle\nu^{II}_{z}\rangle_{z}}\right)^{2}}\cdot\frac{\langle\Delta E^{I}_{z}\Delta E^{II}_{z}\rangle_{z}}{\langle E^{I}_{z}\rangle_{z}\langle E^{II}_{z}\rangle_{z}}+\cdots\,.
\end{align}
As expected, the contribution of the term with $\langle(\Delta E^{I}_{z})^{2}\rangle_{z}/\langle E^{I}_{z}\rangle^{2}_{z}$ is smaller than that with $\langle(\Delta E_{z})^{2}\rangle_{z}/\langle E_{z}\rangle^{2}_{z}$ in Eq. (\ref{MMx}), and even approaches zero when $\langle\nu^{I}_{z}\rangle_{z}\ll \langle\nu^{II}_{z}\rangle_{z}$. Interestingly, the second term on the right side of Eq. (\ref{branch}) explicitly comes from a correlation between $E^{I}_{z}$ and $E^{II}_{z}$. It should be noted that such a correlation between different enzymes does not only affect $\xi$s for specific branching pathways, but also does so for many general situations, because the second and third terms on the right side of Eq. (\ref{MMx}) contain a variation of flux, and flux may change in response to the activity of enzymes or proteins other than only that of an enzyme of interest. Hence, a correlation between different enzymes as well as a variation of each enzyme can play an important role to distinguish a gap between single-cell-level and population-level kinetic constants.

It is also interesting to note that in a limited range our results may be experimentally accessible without measuring metabolites inside a single cell. Briefly, let us consider an enzyme taking as a substrate a molecule which has just passed from the extracellular medium along a membrane transporter like a permease. If $\nu_z \varpropto T_{z}$ where $T_{z}$ denotes transporter level, the second and third terms on the right side of Eq. (\ref{MMx}) can be rewritten as,
\begin{displaymath}
-\frac{\langle S_{z}\rangle_{z}}{K_{M}}\cdot\frac{\langle(\Delta T_{z})^{2}\rangle_{z}}{\langle T_{z}\rangle^{2}_{z}}+\left(1+2\frac{\langle S_{z}\rangle_{z}}{K_{M}}\right)\frac{\langle\Delta E_{z}\Delta T_{z}\rangle_{z}}{\langle E_{z}\rangle_{z}\langle T_{z}\rangle_{z}}\,.
\end{displaymath}
Therefore, the right side of Eq. (\ref{MMx}) includes only the molecular levels measurable by currently available experimental techniques, and can be used to estimate enzyme kinetic parameters for single cells. A more complete description of this idea can be found in the accompanying supplemental material \cite{expsup}.

Even in cases where the primary interest is the kinetics at a population level rather than at a single-cell level, does it matter to consider the effect of cellular variability analyzed so far? Since a specific value of $\xi$ can be changed responding to different conditions rather than kept constant, $\{\widetilde{K}_{i}\}$ will be changed also. This fact indicates that even a functional form of reaction kinetics at a population level will be distorted from its basal form. For example,
the experiments with {\it S. cerevisiae} \cite{rlog2} suggest the relationship $\langle(\Delta E_{z})^{2}\rangle_{z}/\langle E_{z}\rangle^{2}_{z}\approx C_{1}/\langle E_{z}\rangle_{z}+C_{2}$ where $C_{1}$ and $C_{2}$ are constants ($C_{1}\approx 1200$ molecules/cell). By assuming that $\langle(\Delta\nu_{z})^{2}\rangle_{z}$ is sufficiently small, Eqs. (\ref{grd1}) and (\ref{MMx}) lead to
\begin{equation}\label{destroy}
\langle\nu_{z}\rangle_{z}\approx\frac{K_{o}\langle E_{z}\rangle_{z}\langle S_{z}\rangle_{z}}{K_{M}}\left(\frac{1}{1+\langle S_{z}\rangle_{z}/K_{M}}-\frac{C_{1}}{\langle E_{z}\rangle_{z}}-C_{2}\right).
\end{equation}
This result clearly violates the original form of the MM equation, since the MM equation is modified with the introduction of the second and third terms in $(\cdots)$ of Eq. (\ref{destroy}). Although ``effective'' values of kinetic parameters fitted to the original MM equation might work for narrow ranges of $\langle E_{z}\rangle_{z}$ and $\langle S_{z}\rangle_{z}$ that are used for the fitting, it will still be unavoidable to witness the breakdown of the MM kinetics like Eq. (\ref{destroy}) in the face of significant changes of $\langle E_{z}\rangle_{z}$ and $\langle S_{z}\rangle_{z}$ \cite{qbd} conveyed by severe intra- or extracellular condition changes.

So far, we have used the MM equation for $f(E_{z},\{S_{\gamma z}\},\{K_{i}\})$ assuming negligible molecular fluctuations under given $E_{z}$ and $\nu_{z}$. A recent study has suggested that an active transport mechanism of substrates as well as the presence of competitive enzyme inhibitors may manifest the effect of such fluctuations \cite{rfm2}, and this effect can be incorporated in our study through the use of the corresponding $f(E_{z},\{S_{\gamma z}\},\{K_{i}\})$ instead of the MM equation.

In this Letter, we demonstrate that reaction kinetics of a cell population can be systematically deviated from that of single cells by inevitable and significant variations in enzyme abundance. This result would not be only restricted to the case of MM kinetics focused on here; more sophisticated kinetic equations than the MM equation would also face such deviations. Our findings indicate that widely spread discrepancies between {\it in vivo} and {\it in vitro} kinetics might be attributed at least in part to cellular variability, because previously known {\it in vivo} kinetic parameters have been mostly obtained from population-level experiments. We expect that the ultimate development of single-cell metabolomic analysis will greatly facilitate a precise determination of biochemical kinetics. In particular, such single-cell-level analysis can be applied judiciously to key parts of biochemical pathways of which operation is highly sensitive to their kinetic parameters.

The authors thank Elijah Roberts, Thomas Butler, and Kwang-Il Goh for useful discussions. This work was supported by the Institute for Genomic Biology Postdoctoral Fellows Program (P.-J.K.) and an NSF CAREER Award (N.D.P.).


\begin{thebibliography}{99}
\bibitem{ihco}
 For visual clarity, $-O\left\{\frac{\langle(\Delta c_{z})^{2}\rangle_{z}}{\langle\nu_{z}\rangle^{2}_{z}}\right\}+O\left\{\frac{\langle\Delta E_{z}\Delta c_{z}\rangle_{z}}{\langle E_{z}\rangle_{z}\langle\nu_{z}\rangle_{z}}\right\}$ has been omitted at the right side of Eq. (\ref{ih}). Likewise, $-O\left\{\frac{\langle(\Delta E^{II}_{z})^{2}\rangle_{z}}{\langle E^{II}_{z}\rangle^{2}_{z}}\right\}-O\left\{\frac{\langle(\Delta c_{z})^{2}\rangle_{z}}{\langle c_{z}\rangle^{2}_{z}}\right\}+O\left\{\frac{\langle\Delta E^{I}_{z}\Delta c_{z}\rangle_{z}}{\langle E^{I}_{z}\rangle_{z}\langle c_{z}\rangle_{z}}\right\}+O\left\{\frac{\langle\Delta E^{II}_{z}\Delta c_{z}\rangle_{z}}{\langle E^{II}_{z}\rangle_{z}\langle c_{z}\rangle_{z}}\right\}$ has been omitted at the right side of Eq. (\ref{branch}).
\bibitem{qbd}
 Specifically, it occurs when $\Delta\langle E_{z}\rangle_{z}$ and $\Delta\langle S_{z}\rangle_{z}$ cannot satisfy $\left | \frac{\Delta\langle E_{z}\rangle_{z}}{\langle E_{z}\rangle_{z}+C_{2}\langle E_{z}\rangle^{2}_{z}/C_{1}}-\frac{\Delta\langle S_{z}\rangle_{z}}{K_{M}+\langle S_{z}\rangle_{z}}\right |\ll \\ \frac{1}{(1+\langle S_{z}\rangle_{z}/K_{M})(C_{1}/\langle E_{z}\rangle_{z}+C_{2})}-1$.
\end{thebibliography}
\end{document}